# Polarization-independent metasurfaces based on bound states in the continuum with high Q-factor and resonance modulation

XINGYE YANG[1], ALEXANDER ANTONOV[1], ANDREAS AIGNER[1], THOMAS WEBER[1], YOHAN LEE[1], TAO JIANG[1], HAIYANG HU[1] AND ANDREAS TITTL[1, *]

[1]*Chair in Hybrid Nanosystems, Nano-Institute Munich, Faculty of Physics, Ludwig-Maximilians-University Munich, Munich 80539, Germany*
*\*Andreas.Tittl@physik.uni-muenchen.de*

**Abstract:** Metasurfaces offer a powerful platform for effective light manipulation, which is crucial for advanced optical technologies. While designs of polarization-independent structures have reduced the need for polarized illumination, they are often limited by either low Q factors or low resonance modulation. Here, we design and experimentally demonstrate a metasurface with polarization-independent quasi-bound state in the continuum (quasi-BIC), where the unit cell consists of four silicon squares arranged in a two-dimensional array and the resonance properties can be controlled by adjusting the edge length difference between different squares. Our metasurface experimentally achieves a Q factor of approximately 100 and a resonance modulation of around 50%. This work addresses a common limitation in previous designs, which either achieved high Q factors exceeding 200 with a resonance modulation of less than 10%, leading to challenging signal-to-noise ratio requirements, or achieved strong resonance modulation with Q factors of only around 10, limiting light confinement and fine-tuning capabilities. In contrast, our metasurface ensures that the polarization-independent signal is sharp and distinct within the system, reducing the demands on signal-to-noise ratio and improving robustness. Experiments show the consistent performance across different polarization angles. This work contributes to the development of versatile optical devices, enhancing the potential for the practical application of BIC-based designs in areas such as optical filtering and sensing.

## 1. Introduction

Metasurfaces represent a significant step forward in optical manipulation [1, 2]. Traditionally, optical control has relied on bulky lenses and components, however, the advent of metasurfaces has revolutionized this field. With their subwavelength thickness and the use of engineered nanostructures, they allow for precise control over the phase, amplitude, and polarization of light [3, 4]. This transition from conventional to flat optics offers distinct advantages, such as compactness, flexibility, and the potential for on-chip integration, expanding the possibilities of optical devices [5-8].

In the field of metasurfaces, the concept of symmetry-protected bound states in the continuum (BIC) has emerged as a powerful tool for manipulating light [9-12]. They provide a unique mechanism for manipulating light, offering high-quality resonances that can be leveraged in various optical applications such as strong coupling, sensing and optical filtering

[13-18]. However, most quasi-BIC designs and applications are polarization-dependent, which limits their efficiency in applications where the incident beam is unpolarized. For instance, in sensing and filtering scenarios with unpolarized light sources, polarization-dependent designs waste nearly half of the incident light energy or impose strict conditions on the polarization of incident light [17, 19-21]. This inefficiency highlights the need for polarization-independent quasi-BIC designs, which can more effectively harness light across all polarization states, thereby facilitating the transition from academic research to industrial applications.

Currently, there are two prominent approaches to design polarization-independent quasi-BIC resonances. The first approach places the resonators supporting the quasi-BIC mode in a circular arrangement to obtain polarization invariance and large quality factors (Q factors) exceeding 200 [17]. However, these metasurface designs typically exhibit a resonance modulation of only around 10% and face challenges related to their innately low filling factors Consequently, such radial structures can produce low signal-to-noise ratios in sensing experiments, leading to issues for molecular differentiation. Likewise, their low total absorption compared to 2D arrays may limit their efficiency in harnessing light source energy, e.g., for energy conversion applications.

The most widely used approach for obtaining polarization independence in quasi-BIC systems is based on C4 symmetric metasurfaces [19-26]. Such C4 symmetric structures exhibit polarization invariance due to their four-fold rotational symmetry, ensuring identical responses to all incident polarization states. However, so far, many experimental realizations of polarization-independent quasi-BIC metasurfaces have either focused on obtaining large resonances modulations at the cost of Q factor in the terahertz band [27, 28], or have provided generally low resonance modulations, especially when operated in the more fabricationally challenging visible spectrum [19].

In this work, we present an approach based on C4 symmetry for achieving a polarization-independent metasurface by simply altering the length difference of square elements within the unit cell. This alters the structural periodicity and result in Brillouin zone folding, which changes the distribution of photonic modes in the momentum space, converting a BIC into a quasi-BIC at Γ point [19, 29]. In this way, we realize a quasi-BIC metasurface that simultaneously achieves a high experimental Q factor of approximately 100 and a resonance modulation of around 50%. Our design strikes a balance between Q factor and resonance modulation. The high Q factor offers strong field confinement, making it highly sensitive to the surrounding environment. The strong resonance modulation allows the metasurface to respond effectively to external changes like refractive index without being subjected to noise or minor fluctuations. This well-balanced performance suggests that the resonator could be suitable for applications such as sensors, modulators, or filters, where sensitivity and stability against loss or noise are both essential.

## 2. Results

### 2.1 Numerical design

Initially, we identified a suitable structure through simulations that achieves the quasi-BIC resonance within the wavelength range accessible for subsequent experimental characterization using CST Studio Suite (Simulia). A three dimensional (3D) illustration of the metasurface is shown in Figure 1a, where Si-based resonators are positioned on a silica substrate. Each unit cell has periodicities of $P_x=P_y=430$ nm and comprises four rectangular blocks, all of which are square-shaped in the top view (Figure 1b). In the BIC configuration, these blocks have identical side lengths, denoted as $A_1=185$ nm, as indicated by the dashed lines. To achieve the quasi-BIC resonance, we reduce the side lengths of three squares to $A_2$, thereby introducing an in-plane perturbation, leading to a Brillouin zone folding, which transform the nonradiative BIC into a quasi-BIC mode. Meanwhile the whole metasurface retains C4 symmetry, making the resonance polarization-independent. The perturbation parameter dL represents the difference

dL=A1−A2, and its range is set between 0 nm and 30 nm. The resonator height is fixed at 80 nm. Additionally, the scaling factor S is introduced to uniformly adjust the size of the structure in the xy-plane. By introducing dL we change metasurface period from 215 nm to 430 nm. First, We introduce the perturbation of length difference and obtain the quasi-BIC resonance (Figure 1c). Here clearly illustrate the characteristics of the quasi-BIC: as dL increases, more energy couples into the radiation channel, resulting in a broader linewidth. S is set to 0.9. With dL fixed at 20 nm, adjusting Scaling factor (S) allows us for flexible tuning of the resonance position in the spectrum (Figure 1d). These simulations in Figure 1 successfully demonstrate the desired quasi-BIC resonance.

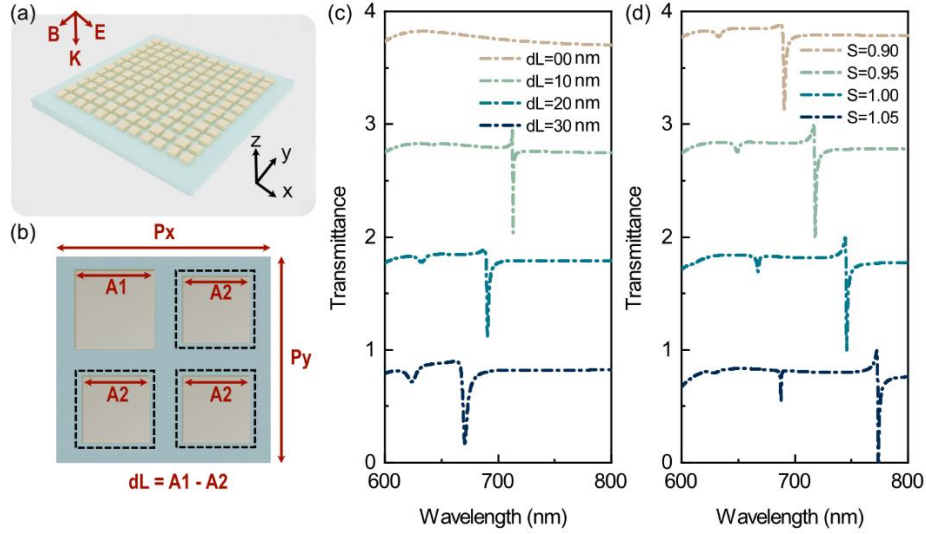

**Fig. 1.** (a) Three-dimensional schematic of the metasurface, showing the Si-based resonators placed on a silica substrate. (b) Top-view of the unit cell structure, with square-shaped elements having dimensions Px=Py=430 nm. The squares are initially identical with side length A1=185 nm. To achieve the quasi-BIC resonance, the side lengths of three squares are reduced to A2, which is smaller than A1. (c) Simulated transmittance spectra of the metasurface for varying dL (the length difference dL=A1−A2), showing the broadening of the resonance peak with increasing dL. (d) Simulated spectra as a function of the scaling factor S with dL=20 nm, demonstrating resonance shift across the spectrum by adjusting the structure size.

Next step is to analyze our quasi-BIC resonance. We clearly observe how BIC (red dashed circle) turns out into quasi-BIC when changing perturbation parameter dL (Figure 2a). Another resonance feature can be observed at a wavelength of 675 nm, however, due to its weak amplitude and spectral separation from the main BIC resonance, we will not focus on this mode in our subsequent analysis. Transmittance spectra for S=1 and dL varying from 0 to 30 nm are simulated (Figure 2a). For the resonance peak with the greater resonance modulation, we extracted the Q factor for each dL using temporal coupled-mode theory (TCMT) [14, 30, 31], as shown in Figure 2b. The corresponding equations are provided in the supplementary materials. The Q factor exhibits an inverse square relationship with the perturbation parameter dL, according to the general rule of quasi-BIC resonance (Figure 2b) [9]. We then selected a representative resonance peak at dL=20 nm for further analysis, performing a multipolar expansion with the incident light polarized in the x-direction (supplementary document Figure S1) [32-35]. The expansion of coupling parameter includes electric dipole, magnetic dipole, electric quadrupole, magnetic quadrupole, and electric octupole moments. By evaluating each term of the expansion we revealed, that the magnetic quadrupole contributes predominantly, which is consistent with the simulation from previous work [20]. Last, electric field distribution in the X-Y-plane and the field enhancement factor, with an enhancement exceeding 30 times, indicating the structure's potential for sensing applications (Figure 2c). The direction of the electric field of the upper and lower parts are in different directions. Due to the introduction of in plane perturbation of the adjacent length difference of the squares, they can no longer compensate each other, resulting in a quasi-BIC resonance accessible from the far field.

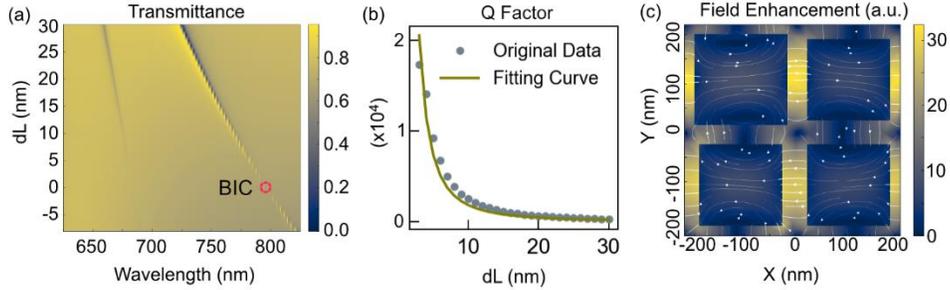

**Fig. 2.** (a) Simulated transmittance spectra for S=1 with dL varying from 0 to 30 nm. (b) Q factor of the resonance peak as a function of dL, showing the characteristic inverse square relationship with dL. (c) Electric field distribution in the xy-plane at the middle cut of the bricks, illustrating the field distribution and enhancement.

*2.2 Experimental realization*

We fabricated the designed nanostructures using a top-down approach with E-beam lithography (for details see Fabrication in supplemental document) and characterized the fabrication results, transmittance spectra, and Q factors. The fabricated structure with dL=30 nm was under SEM imaged, where the expected structure is clearly shown at Figure 3a. Detailed fabrication methods are provided in the supplementary materials. We measured the transmittance spectra for dL ranging from 0 to 30 nm. As the dL increase, the transmittance exhibits a spectral shift similar to that observed in simulations (Figure 1c and Figure 3b). We noticed that, in the experiment, two additional resonances with smaller amplitudes appeared on either side of the resonance peak we focused on. This is caused by a slight deviation of incident angle within our experimental setup, which we will explain in the next section. We focused on the second resonance peak from the right in the spectrum, as its resonance modulation varies significantly with changes in dL.

When the scaling factor is adjusted from 0.9 to 1.05, the structure enlarges, and the transmittance dip red shifts, which also aligns with the simulation trends (Figure 3c). Consequently, we successfully fabricated a metasurface with C4 symmetry, and the experimental results are consistent with the simulations. Additionally, using the TCMT model, we extracted the Q factor and resonance modulation for the resonance peak as dL varied from 10 to 30 nm. Theoretically, the Q factor should become higher at dL=10 nm, but due to influences from fabrication-induced variations in resonator geometry and surface roughness, further enhancement of the Q factor was limited, stabilizing at around 100. At dL=20 nm, the Q factor reached 110, with a resonance modulation exceeding 40%. As dL increased further, the Q factor decreased, consistent with the trend in simulation (Figure 3d).

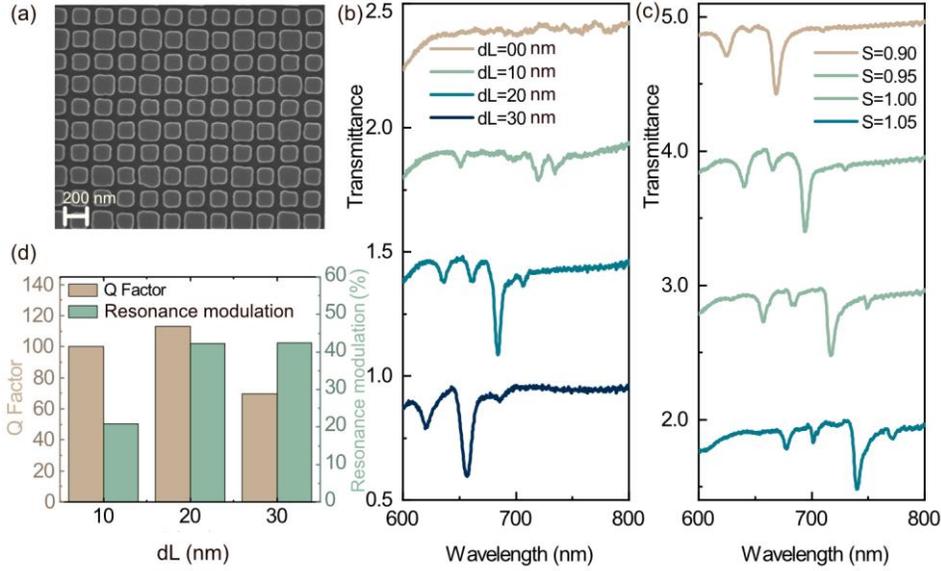

**Fig. 3.** (a) Scanning Electron Microscope (SEM) image of the fabricated metasurface structure with dL=30 nm (b) Experimental transmittance spectra for dL ranging from 0 to 30 nm. (c) Experimental transmittance spectra for varying scaling factors S from 0.9 to 1.05, demonstrating the redshift of the resonance as the structure size increases, consistent with simulation results.

Intriguingly, we observe that in addition to the resonance of interest, two small additional resonance appeared on either side, which were not present in the simulation results. Through simulations, we determined that this was caused by partial oblique incidence of light during the experiment. We simulated transmittance spectra of TE and TM modes, incident at the angles ranging from -5° to +5°. The TM mode under oblique incidence leads to a continuous redshift of the resonance peak at 660 nm (Figure 4b). In the experiment, this continuous shift results in the broadening of the resonance peak. For the TE mode, under oblique incidence, new quasi-BIC modes are introduced off the Γ-point, appearing on either side of the main spectral peak at 740 nm. (Figure 4a) [36, 37]. The combined spectrum of TE and TM modes under oblique incidence cause the mode at 660 nm to broaden and introduce two new quasi-BIC modes around 720 and 760 nm, which are located on either side of the main peak at 740 nm.

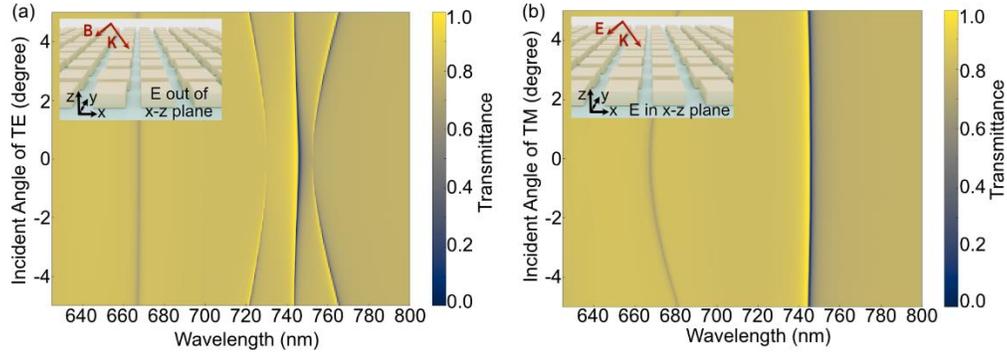

**Fig. 4.** Transmittance spectra of oblique incident TE (a) and TM (b) polarized plane waves for polarization independent quasi-BIC metasurface. The continuous redshift of the resonance peak at 660 nm and its broadening are observed for TM-polarized oblique incident waves. Oblique

incidence of the TE-polarized light leads to revealing two off-Γ quasi-BIC modes on either side of the main peak at 740 nm.

### 2.3 Polarization independence measurements

Following our numerical and experimental analysis, we conducted polarization-dependent measurements. We selected a metasurface with a geometry of S=0.9 and dL=30 nm, corresponding to the top curve in Figure 3c. Alternatively, any other experimental spectrum could have been chosen, this particular one only serves as a demonstration of polarization-independent results. We measured the spectra with the polarization angle ranging from 0° to 90°, in 15° increments, as well as the spectrum without a polarizer. The resulting spectra are shown in Figure 5a. First, we observe that under different polarization angles, the resonance peak remained fixed around 670 nm, with consistent linewidth and resonance modulation, indicating that both the resonance modulation and Q factor remained stable when polarization changes. The resonance frequency shows a slight shift of about 5 nm between 0° and 90°, which is attributed to fabrication imperfections. The Q factors of the measured spectra with different polarization angle are extracted and shown in Figure 5b. We can see a stable Q factor along different polarization angles. Then we extracted the Q factor and resonance modulation with and without the polarizer (Figure 5c). The Q factors range between 70 and 80, with a resonance modulation of approximately 60%. Similarly, depending on specific requirements, we can also select an experimental curve with dL=20 nm, where the Q factor is around 110 and the resonance modulation is approximately 40%. This suggests that our polarization-independent resonance exhibits a high Q factor, indicating good light confinement, and a distinct resonance modulation, offering a better signal-to-noise ratio. Such characteristics could be practically effective in applications like filtering and sensing. Additionally, removing the polarizer allows us to utilize energy from both x- and y-polarized light, thereby improving the ability to efficiently capture the energy incident on the device.

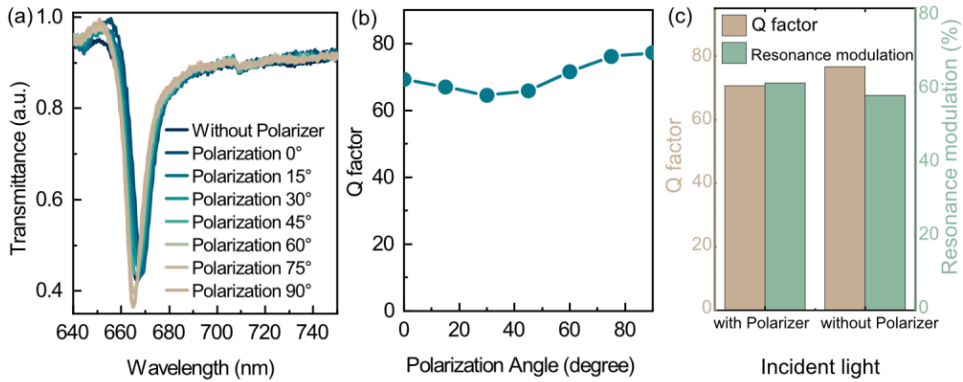

**Fig. 5.** (a) Experimental transmittance spectra of the metasurface, measured for polarization angles from 0° to 90° in 15° increments, along with the spectrum without a polarizer. (b) Q factors extracted from the spectra of different polarization angles. (c) Comparison of Q factor and resonance modulation with and without the polarizer, showing a Q factor between 70 and 80 and a resonance modulation of approximately 60%, highlighting the high Q-factor and strong resonance modulation of the polarization-independent resonance.

### 3. Conclusion

In summary, we have demonstrated a polarization-independent quasi-BIC metasurface based on C4 symmetry, achieved by altering the edge length difference of square elements within the unit cell. The fabricated metasurface exhibits an experimental Q factor of around 100 and a resonance modulation of around 50%. This advantageous combination of large Q factor and strong resonance modulation allows the resonator to effectively enhance and probe processes occurring in the vicinity of the resonators, which is especially important for optical biosensing, where the maximum sensitivity is constrained by the experimental signal-to-noise level. The fabrication and experimental characterization of the metasurface align well with simulation results. The polarization-independent resonance allows for efficient use of light across different polarization states, which could be beneficial in improving capture efficiency when the incident light is unpolarized. Our work advances the development of polarization-independent quasi-BIC metasurfaces mainly towards their ability for applications where both sensitivity and stability are critical, such as in sensors, modulators, or optical filters.

**Funding.** This work was funded by the Deutsche Forschungsgemeinschaft (DFG, German Research Foundation) under grant numbers EXC 2089/1–390776260 and TI 1063/1. We also acknowledge the Bavarian program Solar Energies Go Hybrid (SolTech) and the Center for NanoScience (CeNS). This work was funded by the European Union (ERC, METANEXT, 101078018 and EIC, OMICSENS, 101129734). This work was funded by China Scholarship Council (CSC). Views and opinions expressed are, however, those of the author(s) only and do not necessarily reflect those of the European Union, the European Research Council Executive Agency, or the European Innovation Council and SMEs Executive Agency (EISMEA). Neither the European Union nor the granting authority can be held responsible for them.

**Acknowledgment.** The author would like to thank Jonas Biechteler for his help with E-beam lithography processing, and Chenghao Fan for his support in imaging.

**Disclosures.** The authors declare no conflicts of interest.

**Data availability.** Data underlying the results presented in this paper are available from the corresponding author upon reasonable request.

**Supplemental document.** See supplement document for supporting content.

POLARIZATION-INDEPENDENT METASURFACES BASED ON BOUND STATES IN THE CONTINUUM WITH HIGH Q-FACTOR AND RESONANCE MODULATION: SUPPLEMENTAL DOCUMENT

**Fabrication**

The fabrication process began with the thorough cleaning of fused silica substrates. These substrates were subjected to an ultrasonic bath in acetone, followed by rinsing with isopropanol (IPA). To ensure the substrates were free of contaminants, an additional oxygen plasma etching step was performed. Subsequently, an 80 nm thick layer of amorphous silicon (a-Si) was deposited onto the silica substrate at 250°C using plasma-enhanced chemical vapor deposition (PECVD). A resist layer of PMMA 950k A4 was then applied and baked at 180°C for 3 minutes. To prevent charge accumulation during electron beam lithography (EBL), a conductive polymer (E-Spacer 300Z) was deposited on top of the PMMA layer. The nanostructure pattern was defined in the resist using EBL, operating with an acceleration voltage of 20 kV. After exposure, the resist was developed in a solution of 3:1 isopropanol (IPA) to methyl isobutyl ketone (MIBK) for 2 minutes and 15 seconds. A 50 nm chromium (Cr) layer was subsequently deposited via electron beam evaporation, serving as a hard mask. The metal structures were then lifted off by immersing the substrate in Microposit Remover 1165 at 80°C overnight. The hard mask pattern was transferred into the underlying silicon layer using inductively coupled plasma reactive ion etching (ICP-RIE) with chlorine and fluorine chemistry. The fabrication process was finalized by removing the Cr hard mask through a chromium wet etching.

**Optical Measurement**

The optical characterization of our structures was performed using a commercial white light transmission microscopy setup (WiTec alpha300 series). In this setup, the samples were illuminated with collimated white light from the backside, and the transmitted light was collected using a 50× objective lens (NA = 0.8). During the measurements, a rotating polarizer was used to cover all polarization directions by adjusting it from 0° to 90°.

**TCMT model and resonance depth**

The dynamic equations for the resonance mode with amplitude $a$ are given by[1]:

$$\frac{da}{dt} = (i\omega_0 - \gamma_{\text{tot}})a + \boldsymbol{\kappa}^{\text{T}} \mathbf{s}_+ \tag{S1}$$

$$\mathbf{s}_- = C\mathbf{s}_+ + \boldsymbol{\kappa} a \tag{S2}$$

where $\omega_0$ is the angular frequency of a resonating system. The total damping rate $\gamma_{\text{tot}}$ consists of radiative damping rate $\gamma_{\text{rad}}$ and intrinsic damping rate $\gamma_{\text{int}}$. The resonant mode is excited by the incoming wave $\mathbf{s}_+ = (s_{1+}, s_{2+})^{\text{T}} = (s_{1+}, 0)^{\text{T}}$ from the first port, with the resonant coupling coefficients $\boldsymbol{\kappa} = \left(\sqrt{\gamma_{\text{rad}}}, \sqrt{\gamma_{\text{rad}}}\right)^{\text{T}}$. The outgoing wave is described as $\mathbf{s}_- = (s_{1-}, s_{2-})^{\text{T}}$.

The incoming and outgoing waves at the ports can also couple through a non-resonant scattering pathway, described by a unitary and symmetric matrix $C = e^{j\phi} \begin{pmatrix} r & jt \\ jt & r \end{pmatrix}$ where $\phi$, $r$ and $t$ represent the phase, reflection and transmission constants of direct pathway respectively with $|r|^2 + |t|^2 = 1$.

Assuming a monochromatic time dependence $\frac{da}{dt} = i\omega a$, the solution of equation (S1) is:

$$a = \frac{\sqrt{\gamma_{\text{rad}}} s_{1+}}{i(\omega-\omega_0)+\gamma_{\text{tot}}} \tag{S3}$$

Substituting equation (S1) into equation (S2), the global scattering characteristic of the system can be derived as:

$$\mathbf{s}_- = C\mathbf{s}_+ + \frac{\kappa\kappa^T \mathbf{s}_+}{i(\omega-\omega_0)+\gamma_{\text{tot}}} = \left(C + \frac{\kappa\kappa^T}{i(\omega-\omega_0)+\gamma_{\text{tot}}}\right)\mathbf{s}_+ \equiv S\mathbf{s}_+ \tag{S4}$$

where $S = \begin{pmatrix} s_{11} & s_{12} \\ s_{21} & s_{22} \end{pmatrix} = \left(C + \frac{\kappa\kappa^T}{i(\omega-\omega_0)+\gamma_{\text{tot}}}\right)$ is the scattering matrix for the entire system. The reflectance and transmittance are defined as $R = |s_{11}|^2, T = |s_{21}|^2$, respectively. The transmittance of the quasi BIC system can be expressed by:

$$T = \left| e^{j\phi} jt + \frac{\gamma_{\text{rad}}}{i(\omega-\omega_0)+\gamma_{\text{tot}}} \right|^2 \tag{S5}$$

Then by fitting the transmittance spectra, we can extract the total damping rate $\gamma_{\text{tot}}$. The quality factor $Q$ is calculated by

$$Q = \frac{\omega_0}{2\gamma_{tot}} \tag{S6}$$

Finally, the resonance depth is calculated by

$$\text{Resonance depth} = (1 - \text{minimum of transmittance}) \times 100\% \tag{S7}$$

**Multipole decomposition**

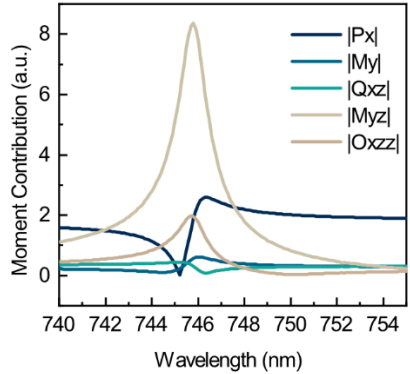

**Fig. S1.** Multipolar expansion analysis of the resonance peak at dL=20 nm for the x-polarized incident light, including contributions from electric dipole (Px), magnetic dipole (My), electric quadrupole (Qxz), magnetic quadrupole (Myz), and electric octupole moments (Oxzz).

The coupling coefficient of the eigenstate with the normally incident light along the *z*-direction with a wave vector $k = \omega/c$ and linear polarization along $\mathbf{e} = \mathbf{e}_x$ can be evaluated as an overlap integral[2, 3]:

$$m_e \propto \int_V \mathbf{J}(\mathbf{r}) \cdot \mathbf{e}\, e^{i\mathbf{k}\cdot\mathbf{r}} dV \propto P_x + \frac{1}{c}M_y - \frac{i\omega}{6c}Q_{xz} - \frac{i\omega^2}{2c}M_{yz} - \frac{\omega^2}{6c^2}O_{xzz} \tag{S8}$$

where the electric dipole, magnetic dipole, electric quadrupole, magnetic quadrupole and electric octupole moments are introduced in a standard way[4]:

$$\mathbf{P} = \frac{i}{\omega}\int_V \mathbf{J}(\mathbf{r})\, dV \tag{S9}$$

$$\mathbf{M} = \frac{1}{2}\int_V \mathbf{r} \times \mathbf{J}(\mathbf{r})\, dV \tag{S10}$$

$$Q_{\alpha\beta} = \frac{3i}{\omega} \int_V \left[ r_\alpha J_\beta(\mathbf{r}) + r_\beta J_\alpha(\mathbf{r}) - \frac{2}{3}\delta_{\alpha\beta} \mathbf{r} \cdot \mathbf{J}(\mathbf{r}) \right] dV \tag{S11}$$

$$M_{\alpha\beta} = \frac{1}{3} \int_V \left[ (\mathbf{r} \times \mathbf{J}(\mathbf{r}))_\alpha r_\beta - r_\alpha (\mathbf{r} \times \mathbf{J}(\mathbf{r}))_\beta \right] dV \tag{S12}$$

$$O_{\alpha\beta\gamma} = \frac{i}{\omega} \left\{ \int_V [J_\alpha r_\beta r_\gamma + r_\alpha J_\beta r_\gamma + r_\alpha r_\beta J_\gamma] dV - \delta_{\alpha\beta} V_\gamma - \delta_{\beta\gamma} V_\alpha - \delta_{\alpha\gamma} V_\beta \right\} \tag{S13}$$

$$V_\alpha = \frac{1}{5} \int_V [2(\mathbf{r} \cdot \mathbf{J}(\mathbf{r})) r_\alpha + \mathbf{r}^2 J_\alpha] dV \tag{S14}$$

We calculate the absolute value of each term in Formula (8) and compare the corresponding contribution of each moment.